\documentclass[referee,a4paper,12pt,traditabstract]{jswsc} 

\usepackage{amsmath}
\usepackage{graphicx}
\usepackage{txfonts}
\usepackage{subfigure}
\usepackage{epstopdf}
\usepackage[displaymath,mathlines]{lineno}
\usepackage[authoryear,round]{natbib}
\usepackage[backref]{hyperref}
\usepackage{url}
\usepackage{xcolor}
\usepackage{booktabs}
\usepackage{multirow} 

\usepackage[normalem]{ulem} 
\hypersetup{colorlinks=true,citecolor=cyan,urlcolor=cyan,linkcolor=blue}

\begin{document}


   \title{Real-time detection of solar flares from ground-based VLF data}

   \authorrunning{P. Teysseyre et al.}

   \author{P. Teysseyre
          \inst{1}
          \and
          C. Briand \inst{1}
          \and
          M. Cohen \inst{2}
          }

   \institute{LIRA, Observatoire de Paris, Université PSL, Sorbonne Université, Université Paris Cité, CY Cergy Paris Université, CNRS, Meudon, France\\
              \email{\href{mailto:pauline.teysseyre@obspm.fr}{pauline.teysseyre@obspm.fr}}
            \and
             Electrical and Computer Engineering (Georgia Tech), Georgia Institute of Technology, Atlanta, GA, USA\\
             }


 
\abstract{ 
A method for real-time solar flare detection and characterization using ground-based Very Low Frequency (VLF, 15 - 45 kHz) data is presented. The D-region, the ionosphere's lowest region, is monitored by VLF waves propagating in the Earth-Ionosphere waveguide. The D-region electron density increases during sudden surges in X-ray radiation from solar flares. This subsequently enhances HF absorption.

By seeking trend changes in VLF phase data, an incremental algorithm finds solar flares. 82.7\% of M and X solar flares are detected within one fourth of their rise time. In addition, several VLF transmitters are monitored simultaneously. Combining information from their phase variations leads to an estimation of the Sun’s X-ray flux. Last, propagation models such as LMP or LWPC are combined with the VLF measurements to compute D-region electron density profiles.

This method and its implementation in a new Python package are a step towards building a more resilient system for flare detection and alerts. Its reliance on ground-based data alone ensures an easy maintenance and a backup in case a satellite failure. It also provides alerts comparable to or faster than those obtained through satellite data, due to shortened data latency.

   }        


   \maketitle

\section{Introduction}\label{sec-intro}

The D-region is the lowest layer of the ionosphere, between 60 and 90 km. It regulates much of the High Frequency (HF, 3 - 30 MHz) wave absorption \citep{zawdie17a}, a frequency range for civil, broadcasting and military use. The altitude of this region makes continuous satellite or balloon monitoring impossible. Instead, it relies on narrow Very Low Frequency/Low Frequency (VLF/LF, 15 - 45 kHz) transmitters, emitting waves propagating in the Earth-ionosphere waveguide. The propagating waves are affected by any changes to the waveguide’s boundaries \citep[e.g.][]{wait_influence_1965, ferguson_effect_1992, thomson_experimental_1993, thomson_lowlatitude_2014, teysseyre2025effect}. Propagation models like LMP \citep{gasdia_new_2021} or LWPC \citep{ferguson_computer_1998} provide modeled VLF amplitude and phase, if a certain ionospheric electron density condition is assumed. Therefore, their comparison with measurements yields estimates of the electron density profiles of the D-region  \citep{thomson_experimental_1993, mcrae2000vlf, thomson_ionospheric_2022, xu2023measurements}. Solar flares produce increases in the Sun’s X-ray flux on timescales from tens of minutes to hours. This energetic radiation increases the ionisation of the neutrals down to the D-region, resulting in an increase of the electron density. HF absorption is stronger when the electron density is higher. Solar flares cause HF absorption to rise on short timescales (ten minutes to an hour). Fast and accurate real-time detection of solar flares is therefore crucial in mitigating their impact on radio operation. \\

Different flare detection systems already exist. Most issued alerts are based on solar X-ray flux measured by the Geostationary Operational Environmental Satellite (GOES) from the National Oceanic and Atmospheric Administration (NOAA). The NOAA’s Space Weather Prediction Center (SWPC) provides both post-flare reports and alerts when the solar X-ray flux attains an M5 ($5 \times 10^{-5}$ W.m$^{-2}$) threshold \citep{rockville2019customer}. The reliability of GOES data and alerts stems from the satellites’ continuous operation and direct measurement of the Sun’s X-ray flux. It also presents a latency of four to six minutes \citep{george2019developing, hudson2025anticipating}, which is significant compared to solar flare onset timescales. Solar flares also cause UV and EUV emissions to rise, along with the Sun’s soft X-ray flux. Therefore, alternative alert systems use different wavelengths \citep[e.g.][]{bonte2013sofast, kraaikamp2015solar, lu2024automatic}.  Those alerts detect 80 to 88\% of M and X-class flares, 5 minutes before the peak flux on average \citep{kraaikamp2015solar}. While those systems present robust algorithms allowing the detection of most flares, they rely on satellite data. Flare detection is made more complex in particular cases (e.g. the South Atlantic Anomaly; \citep{bonte2013sofast}) and could be influenced by eruptive events like filaments \citep{lu2024automatic}. Neural network methods \citep{fernandez2002automatic, qu2003automatic} were also investigated, with improved detection rates and capabilities for real-time processing. However, those attempts similarly relied on satellite data.\\

Developing an alert system based on ground-based data presents several advantages. First, ground-based instruments are less vulnerable to strong space-weather conditions than satellites, which can suffer from electronic glitches, loss of position and communication. Building a resilient flare detection service thus involves collecting data from various sources. It also requires relying on diverse data analysis methods to act as backups should one fail. VLF in particular show potential for real-time flare alerts. GOES’s direct measurements of solar X-ray flux are invaluable, but they saturate at X17, contrary to VLF \citep{thomson2005large}. VLF measurements also offer a shorter data latency (less than one minute, compared to four to six minutes for satellites). The receiver's low cost enables the deployment of worldwide networks. Note also that the VLF instruments allow the detection of Transient Luminous Events \citep[e.g.][]{mika_vlf_2008} and electron precipitations from the magnetosphere \citep[e.g.][]{potemra1973, clilverd2009}, two other forcing sources of the D-region electron density. The aim is to maximise both the probability of a flare detection and the time between the detection and the flare peak. Two ground-based VLF alert systems were thus suggested. The first one relied on the Global Ionospheric Flare Detection System (GIFDS) network of VLF electric antennas \citep{wenzel2016global, banys2017propagation}. This method depends on the capability of determining a quiet-day profile relying on the previous days, which may be an issue in active periods. \cite{george2019developing} suggested another flare nowcasting approach. An estimate of the Sun's X-ray flux was provided from a linear regression of VLF phase data and additional parameters such as the solar zenith angle. However, the authors noted that this idea presented two limitations: the need to compute the start of the solar flare, and the reliance on near real-time GOES measurements. To the best of our knowledge, none of these methods are yet implemented for real-time applications. We therefore developed a real-time flare detection system based on VLF measurements alone, with the following aims:
\begin{enumerate}
   \item Detect the occurrence of solar flares very early after its onset, and more crucially before its peak;
   \item Estimate the Sun's X-ray flux;
   \item Estimate the duration of solar flares;
   \item Estimate the HF absorption resulting from the flare and the impacted regions.
\end{enumerate}

Here is presented methods and a tool to address the first two points. The latter two will be implemented in a future version of this tool. The preparation of the data (amplitude calibration and phase correction) is explained in Section \ref{sec-datacor}. Section \ref{sec-detection} details the flare detection method based on breakpoint detection and its performance. Section \ref{sec-fluxestimation} addresses the question of solar flux estimation using a probabilistic approach. Section \ref{sec-necomp} presents the method for real-time estimation of the electron density in the D-region. Finally, Section \ref{sec-package} presents the Python package performing these different steps. In particular, a general workflow is presented, summarising the various steps described previously. Section \ref{sec-improving} details some of the possibilities for improvement of the methods' performances.

\section{Instrumentation and calibration}\label{sec-datacor}

\begin{table}
\begin{center}
\begin{tabular}{c|c|c|c|c|c|c}
   \bf{Call-sign} & \bf{Frequency (kHz)} & \bf{Latitude (°)} & \bf{Longitude (°)} & \textbf{Country} & \textbf{Distance (km)} & \textbf{$\delta f$ (Hz)}\\
   GQD & 22.1 & +54.7 & -2.9 & UK & 921 & 0\\
   NAA & 24.0 & +44.6 & -67.3 & USA & 5206 & 2.8572 $\times 10^{-5}$ \\
   NRK & 37.5 & +63.9 & -22.5 & Iceland & 2372 & 1.4136 $\times 10^{-4}$\\
   NSY & 45.9 & +37.1 & +14.4 & Italy & 1519 & 2.0697 $\times 10^{-4}$\\
\end{tabular}
\end{center}
\caption{VLF transmitters used in Nançay for the flare detection and their characteristics. The sixth column indicates the transmitter-receiver distance.} The $\delta f$ parameter is introduced in Section \ref{sec-datacor}
\label{tab-stations}
\end{table}

Amplitude calibration and phase correction constitute the initial step in analysis. Amplitude calibration involves converting voltage into magnetic field strength. This study will examine data from an AWESOME receiver \citep{cohen_broadband_2018} situated in Nançay, Sologne, France. The AWESOME instrument consists of two orthogonal air-core loop antennas. In Nançay, those loops are triangles with a 4.47 m base and a 2.28 m height, and made of 11 turns of copper wire. It is thus sensitive to magnetic fields above 357~fT.$\sqrt{Hz}$, with absolute timing accuracy of 15-20~ns and 96~dB of dynamic range. It records broadband (waveforms) and narrowband measurements which are used here. Narrowband data consist of amplitude and phase data within a 200~Hz bandwidth centred around a transmitter frequency, at a 1~Hz sampling rate. Four transmitters are used for the real-time analysis (Table \ref{tab-stations}). \\

The instrument’s response to a known signal is used to calculate the amplitude calibration factors through an in-built amplitude-calibration device. VLF transmitters encode communication signals typically with minimum shift keying (MSK) which alternates between one of two frequencies to communicate bits. The phase is obtained from the real-time demodulation of the signal \citep{gross2018polarization}. Phase correction involves unwrapping, then detrending the raw phase. Unwrapping compensates for spurious jumps introduced by constraining the phase between -180° and 180°. The phase presents an artificial slope coming from a small drift of the transmitter's frequency with time. Post-processing analysis compensates for this slope by fitting a straight line to the data and eliminating the trend. Yet, this relies on previous data, which may not be immediately available if the VLF transmission is disrupted. Instead, another method is employed for real-time processing. The total phase is written as:

\begin{equation}\label{eq-phasedrift}
   \phi = \phi_0 + \delta f \frac{2 \pi}{360} t
\end{equation}

where $\phi$ is the unwrapped phase measured and $\phi_0$ is the stable phase. $\delta f$ is the small shift in frequency causing the phase drift, and $t$ is the time (in s) since the start of the day. Given $\delta f$, $\phi_0$ is derived from $\phi$ with no need for past data. $\delta f$ depends on the transmitter and varies along the year, following adjustments from the operators \citep{gross2018polarization, cannon2025avid}. It is approximated by unwrapping the phase on consecutive days and minimising the phase jumps between days. The values of $\delta f$ for the transmitters monitored at Nançay for the period from November 2023 to December 2024 are presented in Table \ref{tab-stations}. The users are invited to check those values monthly, as they may vary in time. Once the two-step correction is complete, the phase trend is accurate, though its absolute value is still determined by the first point of each day. As the subsequent analysis relies exclusively on phase variations, absolute phase calibration is not required.

To further mitigate noise, a moving median filter is applied to the VLF phase and amplitude, with a time-resolution $t_r$. In addition, a flag system is implemented to highlight usable periods. Flags denote intervals during which a transmitter is inactive, identified by a low amplitude signal that varies according to environmental noise levels. In Nançay, this noise is estimated to be 0.5 pT. Flags also eliminate nighttime periods, defined by the solar zenith angle value: nighttime starts when the solar zenith angle is above 85° at either the receiver or the transmitter. This value of 85° corresponds to the minima in the VLF data due to sunrise or sunset. This is a conservative value that satisfies all seasons. Since we consider the two extremes of the path, we are not constrained by its length or obliquity. \\

\section{Flare detection}\label{sec-detection}

After applying corrections and flags, solar flares are identified. The phase always shows a peak during a flare, while amplitude can present successions of peaks and dips, depending on the distance between the transmitter and receiver \citep{Briand22}. Thus, the detection is based on phase data, as it is better correlated with the solar X-ray flux \citep{george2019developing}.  In post-processing, flares are identified through phase peaks and high solar X-ray flux. Our goal is to detect flares before their peak. The solution is to seek changes in the phase trend (\textit{breakpoints} hereafter). The primary challenge in breakpoint detection lies in optimising the trade-off for detection sensitivity. A comprehensive detection of all breakpoints inherently risks the inclusion of spurious signals arising from instrumental or environmental noise. In contrast, focusing solely on large trend changes misses smaller events. Another limitation arises from the need for fast, low-delay computation. The incremental algorithm meets those criteria \citep{guralnik1999} and is described in Section \ref{sec-detectionmethod}. \\

\begin{figure}
    \centering
    \includegraphics[width=\columnwidth]{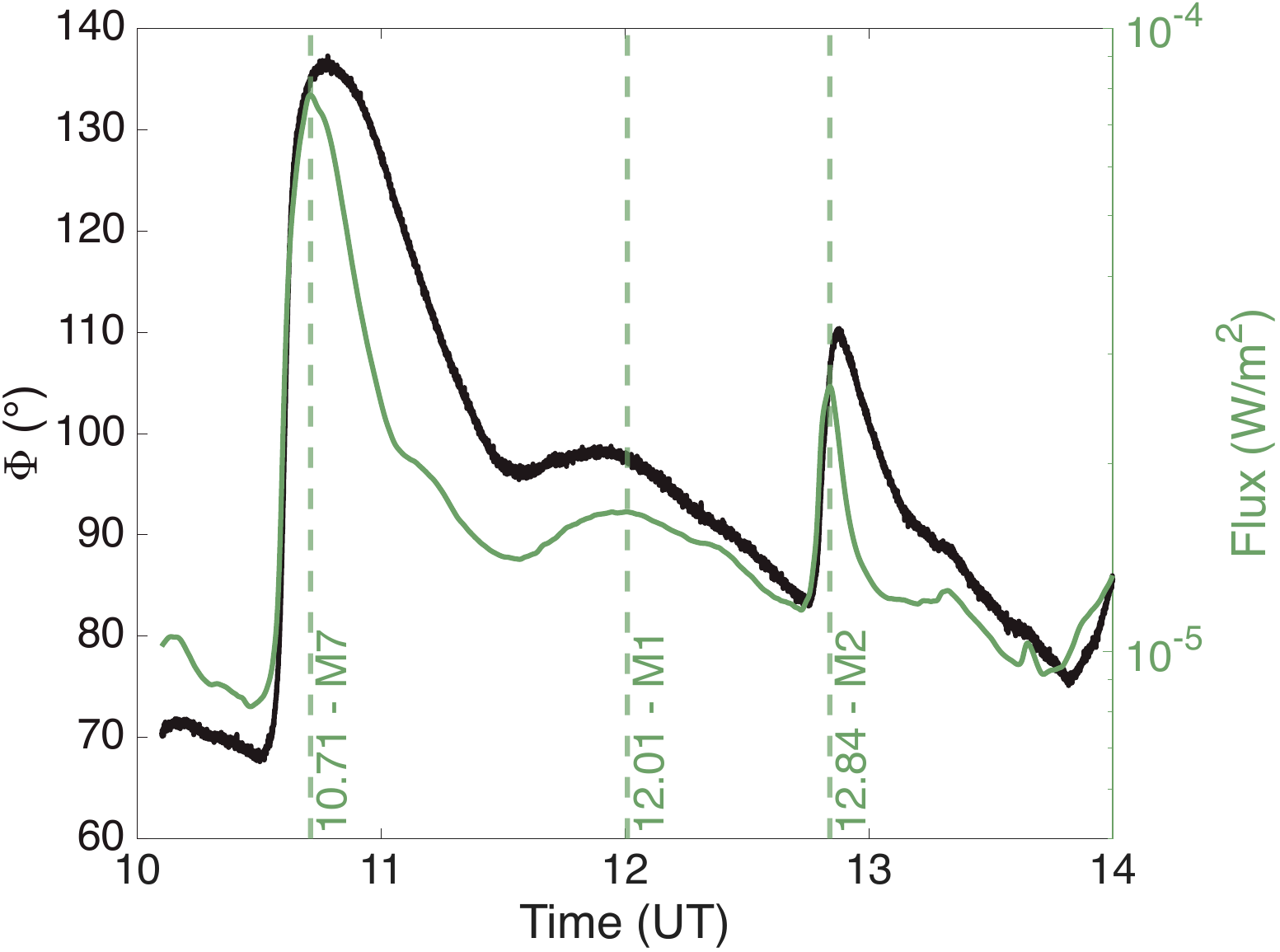}
    \caption{Phase ($\Phi$, in °, black curve) and soft X-ray flux (in W/m$^2$, green curve) on 2024/07/28. Three flares are presented and taken as illustration for the reminder of the paper. They are denoted by their soft X-ray flux peak time: 10.71, 12.01 and 12.84 UT.}
    \label{fig-firstex}
\end{figure}

To test this method, a full year of data was processed, covering the period from 2023/11/01 to 2024/10/31. This period includes highly active months like May 2024 (which included numerous X-flares), as well as quieter periods such as November and December 2023.  Additionally, this period includes instances when the transmitters were inactive, likely due to maintenance. To give more specific illustrations, the example of three consecutive flares on 2024/07/28, is taken hereafter. The NRK phase and the X-ray flux defining those flares are displayed in Figure \ref{fig-firstex}. Hereafter, GOES X-ray flux will always refer to the soft X-ray band, even though hard X-ray are more crucial for strong flares \citep{Briand22}.

\subsection{The incremental algorithm}\label{sec-detectionmethod}

Breakpoints are detected through the incremental algorithm. This method minimises the distance of the data points to a fit. We assume that a straight line correctly fits the phase, as it is constant during quiet periods. This approach ensures the fast computation required for real-time treatment, compared to using a more complex fit. This distance to minimise is measured by the likelihood criteria $L$:
\begin{equation}
   L = m \log{\frac{R}{m}},
\end{equation}
with $m$ the number of data points, and $R$ the sum of residues of data points to the linear fit. It is assumed that different segments present different variances (i.e. heteroscedastic hypothesis. This was checked on data from 2024/07/28, where the variance of different data segments varies by about 30\%. \\

The first data point is automatically declared to be a breakpoint. Afterwards, each new data point is a potential breakpoint. For each one, two hypotheses are examined:
\begin{enumerate}
   \item It follows the previous phase trend. The likelihood criterion is then computed and denoted $L_{no change}$, representing the distance to the fit over the entire dataset. 
   \item This data point introduces a break in the phase trend. It is further assumed that this break occurs five data points prior, as five points are needed to accurately fit a straight line. This assumption ensures fast detection while guaranteeing that the new trend is accurate. Moreover, it limits the computation time arising from checking each data point. Last, the new global likelihood criterion is calculated by adding the local likelihood criteria from both segments (separated by the breakpoint) and is denoted $L_{change}$.
   \end{enumerate}
   $\delta$ is a user-defined threshold; the new breakpoint is kept when it satisfies
   \begin{equation}\label{eq-delta}
   \frac{(L_{change} - L_{no change})}{L_{no change}} > \delta
   \end{equation}
   Then, the phase slope associated with the breakpoint is named $s$ and is used to determine the state of the flare evolution (see Section \ref{sec-anameth})

\subsection{Parameter tuning for the detection method}\label{sec-detectionres}

\begin{figure}
    \centering
    \includegraphics[width=\linewidth]{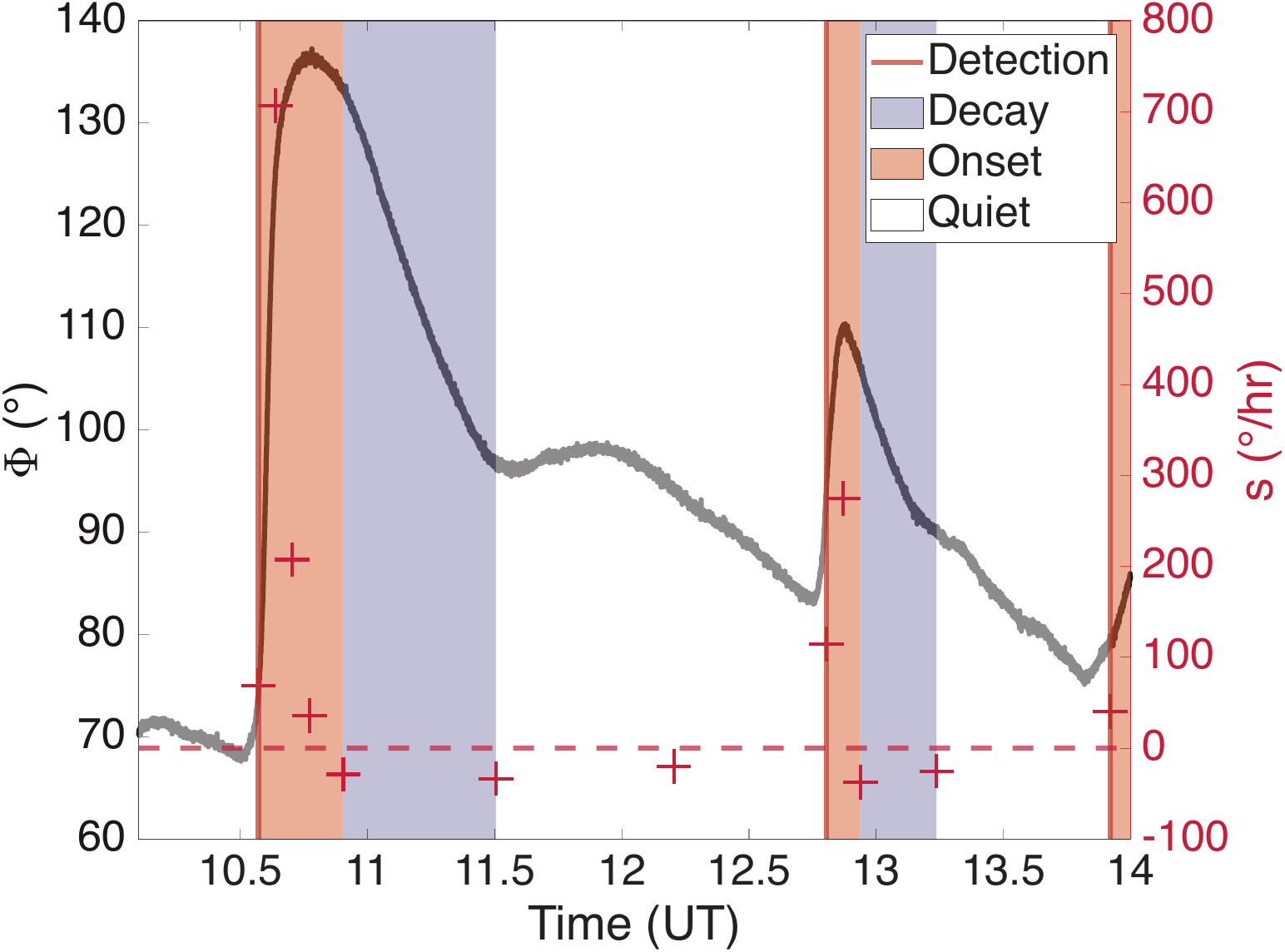}
    \caption{NRK phase (black line, also shown in Fig. \ref{fig-firstex}) and detected breakpoints (red crosses) for 2024/07/28. The right vertical axis gives the scale of the slope. Four periods are indicated: flare detection (red vertical lines), flare onset (orange area), flare decay (light purple) and quiet (white). The breakpoint classification is detailed in Section \ref{sec-anameth}.} 
    \label{fig-exanabp}
\end{figure}

The detection performance is controlled by two parameters: $\delta$ and $t_r$ (Equation \ref{eq-delta} and Section \ref{sec-datacor} respectively). The goal is to find the optimal values to detect the maximum of flares and minimize the false detections. The optimal values of $\delta$ and $t_r$ were estimated by analysing data from the NRK transmitter (Table \ref{tab-stations}) over one-year period of reference. During this period, 390 M and 17 X-flares (407 total) occurred when the transmitter was on and the path remained entirely within daylight. Each flare above C3 was initially found using GOES X-ray flux measurements. Periods when NRK was off were identified during the flagging step (Section \ref{sec-datacor}). Of the 407 flares, 271 have a noticeable response in the data. The others were too close to nighttime or to another more powerful flare. All X-flares were observed in the VLF phase measurement. Figure \ref{fig-exanabp} shows an example of breakpoint detection from phase measurement of NRK performed on 2024/07/28. For each flare, a visual inspection confirmed the identification. In parallel, the phase maximum time is determined. A \textit{detection threshold} is also defined as the 75th quantile of quiet-time slopes. A flare is `detected' when the breakpoint's slope exceeds this detection threshold.  \\

\begin{figure}
    \centering
    \includegraphics[width=\linewidth]{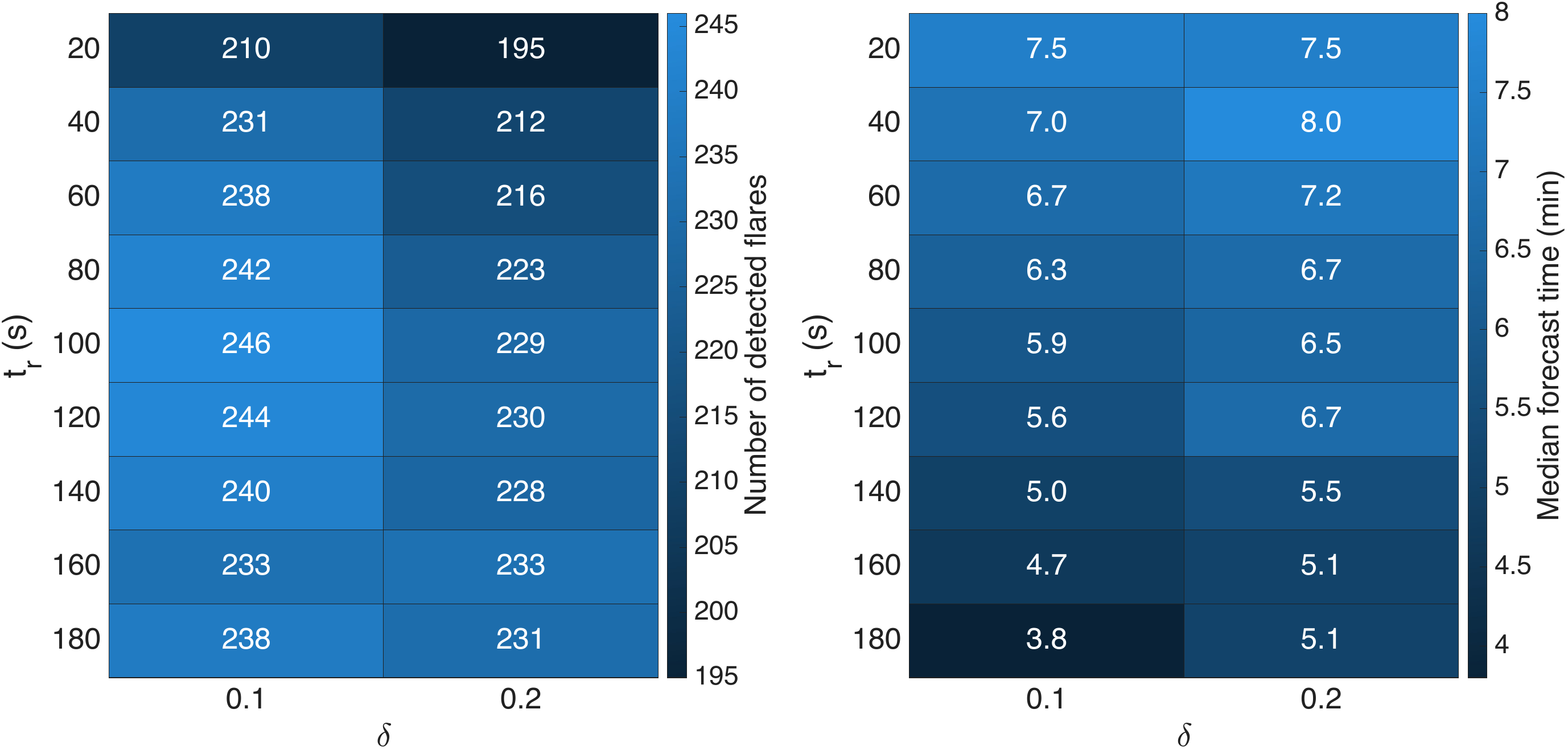}
    \caption{Number of flares for which a positive phase slope was detected (left) and median time delay (right) of the detection method applied to one year of phase data from NRK for different combinations of $\delta$ and $t_r$.}
    \label{fig-resparams}
\end{figure}

The forecast time is defined as the time difference between the flare detection and phase maximum. Optimal parameter values maximize flare detection and forecast time, while minimizing computation time. The forecast time is flare-dependent: the mean forecast time is presented in Figure \ref{fig-resparams}. Detecting more flares was prioritized over a longer forecast time. $\delta = 0.1$ ensures the maximum number of detected flare: this parameter value improves the number of detected flare by about 20. $t_r$ was then determined to ensure a balance between this number and the forecast time. $t_r$ is thus set to 60~s, a rate that optimizes the code running-time and makes it more manageable. With these parameters, 82.7\% M and X flares visible in the training set were detected. All X-flares were detected. 24 of the 47 undetected flares occurred during the decay of a nearby flare. The other undetected flares were 8 faint flares (the phase increase $\Delta \Phi$ is below 10°), one short-duration flare (less than 3 min), and 14 flares presenting a positive slope below the detection threshold. The optimal parameters values were determined from NRK data. However, they depend on both environmental noise and the difference in slope between flares and quiet times. Thus, they vary for each transmitter-receiver path. Nevertheless, as long as $\delta < 0.2$ and $t_r < 120 s$, the detection performance are comparable (Fig. \ref{fig-resparams}). These parameter values should thus hold for any transmitter-receiver path without high noise levels or weak phase slopes during flares. \\

\begin{figure}
    \centering
    \includegraphics[width=\linewidth]{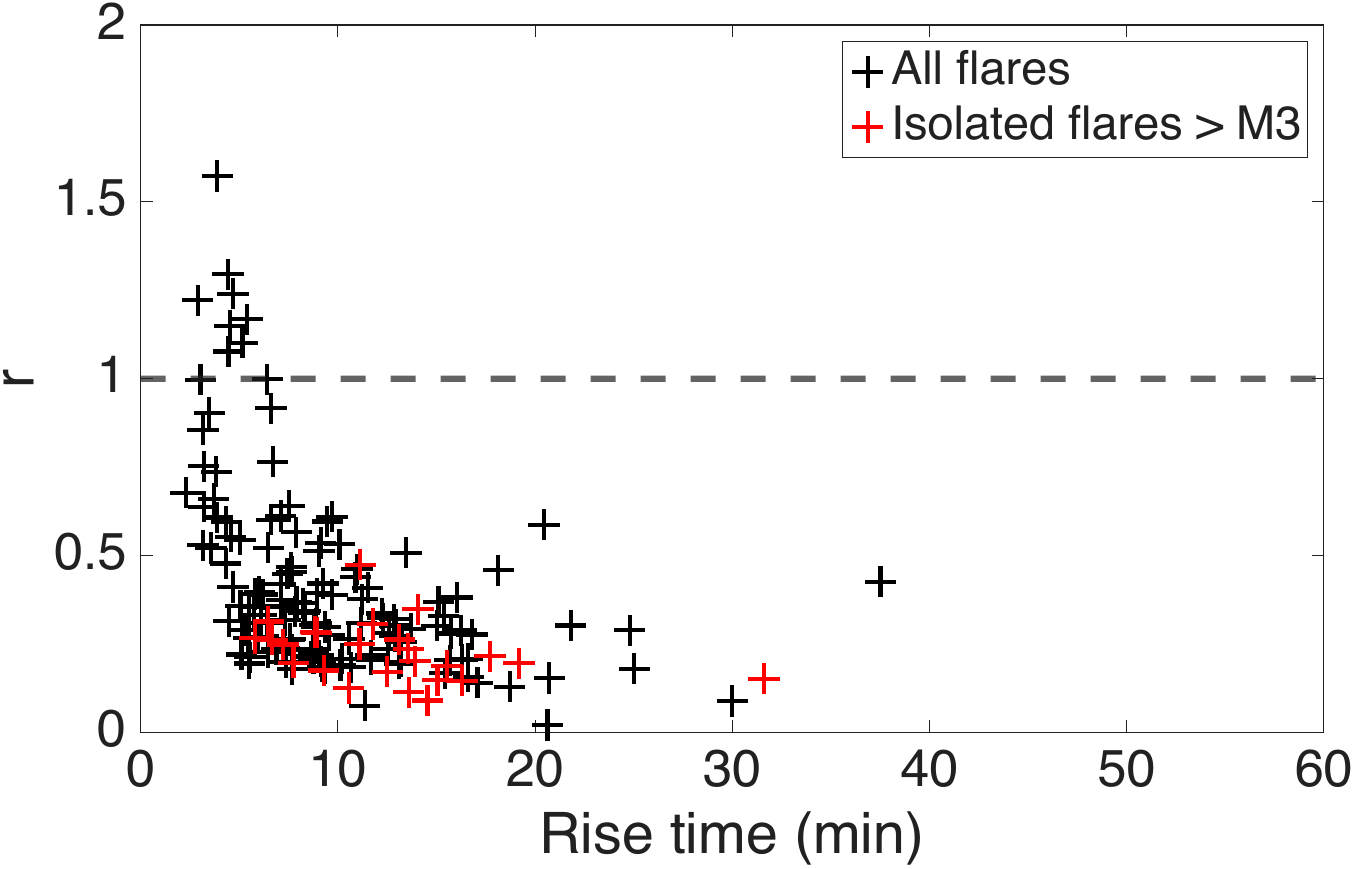}
    \caption{Ratio between the detection time and the phase rise time ($r$) vs phase rise time (in minutes).}
    \label{fig-delay}
\end{figure}

The goal is to reduce the detection time, that is the time between the flare start (when the VLF-phase slopes become positive) and the flare detection. The detection time depends heavily on the rise time (the time between the flare start and the phase maximum). A ratio of detection time to rise time greater than one indicates that the flare was detection after the phase reaches its peaks (Figure \ref{fig-delay}). Most of the flares are characterised by a rise time shorter than 6.5 minutes. If the flares are constrained to isolated flares above M3, the ratio presents a median of 0.22 and a mean of 0.24 (red crosses in Figure \ref{fig-delay}). Strong isolated flares are thus detected at one fourth of their rise time on average. Also, as anticipated, a longer rise time leads to an earlier detection. \\

To quantify the quality of the detection, we check the delay between the soft X-ray flux increase and the detection. The ionosphere has a slight lag in its response, as reflected in the time-delay from amplitude measurements \citep{appleton1953note, basak2013effective, Briand22}. The VLF-flare detection is on average 7.1 minutes after the start of an X-ray flux increase. However, the average time from VLF phase increase to flare detection is 2.6 minutes. For comparison, GOES alerts are triggered by M5 threshold with a latency of 4 to 6 minutes \citep{george2019developing, hudson2025anticipating}. Our method relies on the detection of the flare start for any flare strength. \\

\subsection{Breakpoint analysis}\label{sec-anameth}

At this stage, a breakpoint has been detected, but not classified. Therefore, each new breakpoint must be analysed to determine whether it is quiet time, flare onset, flare maximum, or flare decay.  \\

The first step of this classification is based on the computation $\Delta \Phi$, the variations in phase during a solar flare. It only depends on the previous measured breakpoints. $\Delta \Phi$ is calculated through:

\begin{equation}\label{eq-DP}
    \Delta \Phi = \Phi - \left(\Phi_q + s \times (t - t_q)\right)
\end{equation}

\begin{figure}
   \includegraphics[width=\linewidth]{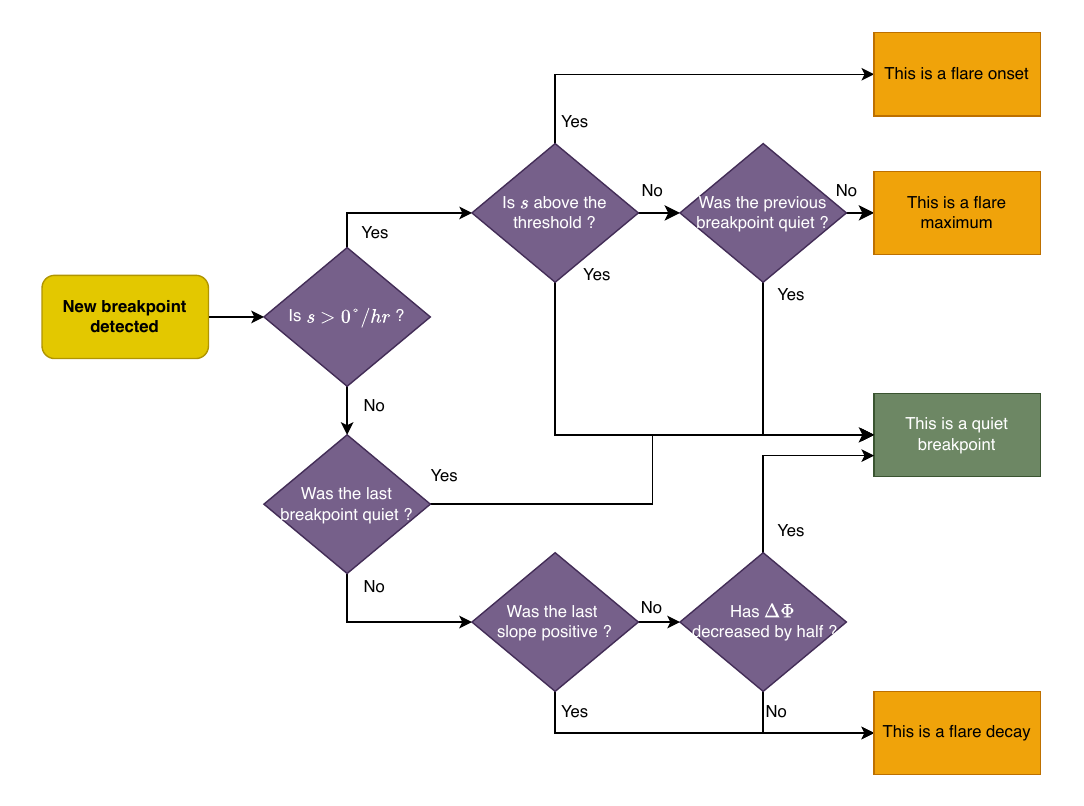}
   \caption{Breakpoint identification scheme. It starts from a "New breakpoint detection" (left side), and continue with the slope $s$ sign, which will determine whether the breakpoint is due to a flare onset, flare maximum, flare decay or is still in quiescent periods.}   
    \label{fig-anabp}
\end{figure}

$\Phi_q$, $s$ and $t_q$ are respectively the phase measured at the last breakpoint, the slope, and time of the last breakpoint. $\Phi$ and $t$ are the current phase and time. The breakpoint is then classified using the algorithm presented in Figure \ref{fig-anabp}. \\

After building a catalogue of all breakpoints from the one-year period described previously, they are sorted by ‘flare’ and ‘quiet period’. Comparison with GOES-X catalogue shows that 94\% of our identification was correct. Let's call `positive' and `negative' times categorised as flare or quiet from our identification. `True' and `false' refer to the correctness of our identification. 'False' means that the identification is wrong. Thus, a `true positive' denotes a flare that has been correctly detected, while a `false negative' represents a flare identified as a quiet period. Any undetected C flare was considered as quiet time, as those faint flares will not greatly impact the ionosphere; but detected C flares were considered `true positives' since they induced an actual change in the phase data. Our identification includes 51\% for true positives and 43\% of true negatives. The incorrect identifications consist of 2\% of false negatives and 4\% of false positives. Some false negatives were due to flares producing a weak phase slope, and false positives were occasionally present during quiet periods. \\

Figure \ref{fig-exanabp} presents the different detected breakpoints and their classifications on the 2024/07/28. The flare peaking at 10.71 UT (resp. 12.84 UT) is detected 12.6 minutes (resp. 4.1 minutes) before the phase peak, and 8.4 min (resp. 2.1 min) before the X-ray flux peak. The flare at 12.01 UT remains undetected, as the flare at 10.71 UT did not recovered its quiescent level when it occurred. Each period (onset, quiet and decay) is delimited by a breakpoint. Thus, the onset period of the flare is overestimated for both the M7 and M2 flare. \\

\section{Solar flux estimation}\label{sec-fluxestimation}
\subsection{Method}\label{sec-fluxest}
Once a breakpoint has been classified as a flare the solar X-ray flux is estimated from $\Delta \Phi$. Daily variability in the ionosphere depends on physical parameters (such as the X-ray spectrum, the solar zenith angle or the season). Thus, there is no bijection between $\Delta \Phi$ and the X-ray flux. Instead, there is a dispersion of $\Delta \Phi$ regarding the solar X-ray flux \citep[Figure \ref{fig-fluxNAA}, top panel and ][]{thomson2005large}. This implies that a probabilistic approach must be employed instead of a direct derivation of the flux from the phase variation. \\

The probability P of the flux value being $F$ is computed from Bayesian statistics, given $\Delta \Phi_1$ and $\Delta \Phi_2$ the phase variation for two transmitters (as measured with a single receiver), and background information $I$:
\begin{align}
   P(F|\Delta \Phi_1, \Delta \Phi_2, I)  &\propto  P(\Delta \Phi_2, \Delta \Phi_1 | F, I) \times P(F | I) \nonumber \\
   & \propto P(\Delta \Phi_1 | F, I) \times P(\Delta \Phi_2 | F, I) \times P(F | I) \nonumber \\
   & \propto \frac{P(F |\Delta \Phi_1, I)}{P(F | I)} \frac{P(F |\Delta \Phi_2, I)}{P(F | I)} \times P(F | I) \label{eq-proba}
\end{align}

$P(F|\Delta \Phi_1, \Delta \Phi_2, I)$ is called the posterior density function, and represents the probability of the X-ray flux being $F$, knowing $\Delta \Phi_1$ and $\Delta \Phi_2$. Statistical independence between $\Delta \Phi_1$ and $\Delta \Phi_2$ is assumed to simplify this computation, and probabilities independent of $F$ are constants. This is justified, since the variations of $\Delta \Phi$ depend on the interactions of the different propagating modes in the Earth-Ionosphere waveguide, which is notably parametrised by the transmitter frequency and ground conductivity. The phase variations also depend on the solar zenith angle. In most cases, the transmitter-receiver paths are far enough that the solar zenith angles and ground conductivities are distinct (except for the GQD and GVT transmitters). This simplification can be checked with the $\Delta \Phi$ data for various transmitters. However, the limited number of flares in the data set, particularly X-class ones, would make this calculation unreliable at the moment. This option remains available for later execution, pending further data acquisition. \\

Equation \ref{eq-proba} can be generalised to any number of transmitters. Therefore, the X-ray flux is estimated from the prior $P(F|I)$ and the successive conditional probability density functions associated with each $\Delta \Phi_i$. Since the prior represents the probability density of the flux without knowledge of the $\Delta \Phi$ values, it is taken as uniform from $10^{-6}$ to $10^{-3}$ W.m$^{-2}$. 

\subsection{PDFs from a single transmitter}

To compute the PDF in Equation \ref{eq-proba}, it is necessary to discretise the values of $\Delta \Phi$ and the X-ray flux. $\Delta \Phi$ is computed from fully resolved M and X flares following Eq. \ref{eq-DP}, and discretized into five bins. If $A$ is the maximum of $\Delta \Phi$, then the edges of these bins are 0, $A$/5, $A$/2, $2\times$ $A$/3, $4\times$ $A$/5 and $A$. $A$ depends on each transmitter-receiver path. As $A$ depends on each propagation path, the $\Delta \Phi$ binning will depend on each transmitter and receiver. The X-ray flux values are similarly discretized, with bin edges M, M3, M6, M9, X2, X5 and X10. The choice of the bin edges is motivated by trying to limit the impact of the overabundance of weak M flares compared to the stronger X flares, by having larger bins for low $\Delta \Phi$ values. However, it is also important to distinguish between M1 and M5 or M9 flares, especially since NOAA’s Space Weather scales for Radio Blackout change at M5 and X1.\\ 

The probability density functions thus computed are then fitted by normalised Gaussian distributions to smooth them. The normalisation first ensures that the density functions never go to zero, leaving a (small) probability that the flux differs from the best value computed from a single transmitter. It also facilitates the storage of these probability functions, since only the average and standard error are needed for each $\Delta \Phi$ bin. \\

\begin{figure}
    \centering
    \includegraphics[width=\linewidth]{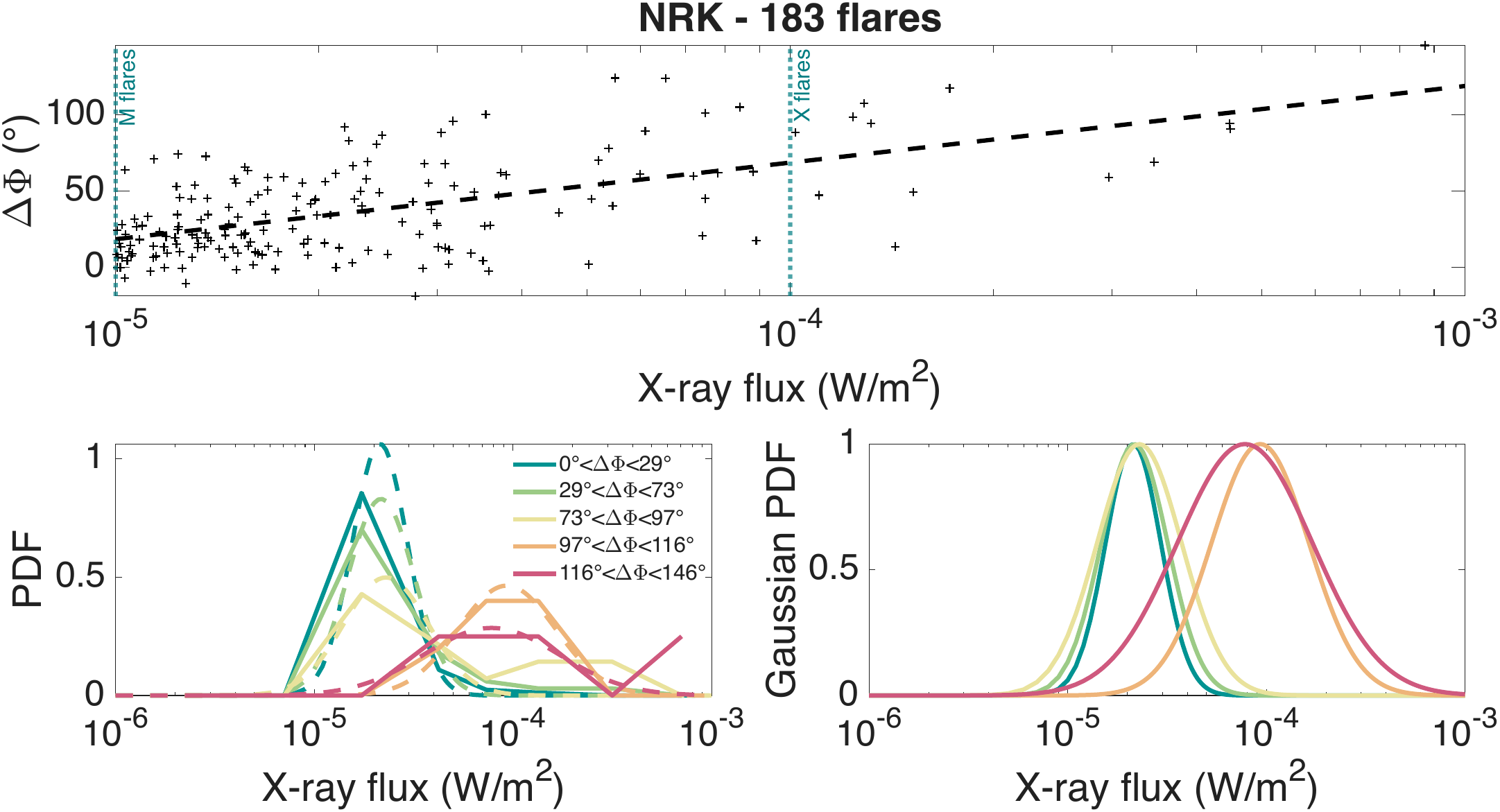}
    \caption{Probability density functions (PDF) for NRK. Top panel: $\Delta \Phi$ vs. GOES X-ray flux. Bottom left panel: PDFs for discretized phase. Bottom right: PDF fitted by a normalized Gaussian distribution.}
    \label{fig-fluxNAA}
\end{figure}

Figure \ref{fig-fluxNAA} shows the application of the method to 183 isolated flares observed from NRK. Phase variations between 0 and 97° can hardly distinguish between M-flare strength. Furthermore, since the training only contained 13 X-class flares, the PDF are not well-prepared for very strong flare. We will discuss these limitations in Section \ref{sec-discussion}.

\begin{figure}
    \centering
    \includegraphics[width=\linewidth]{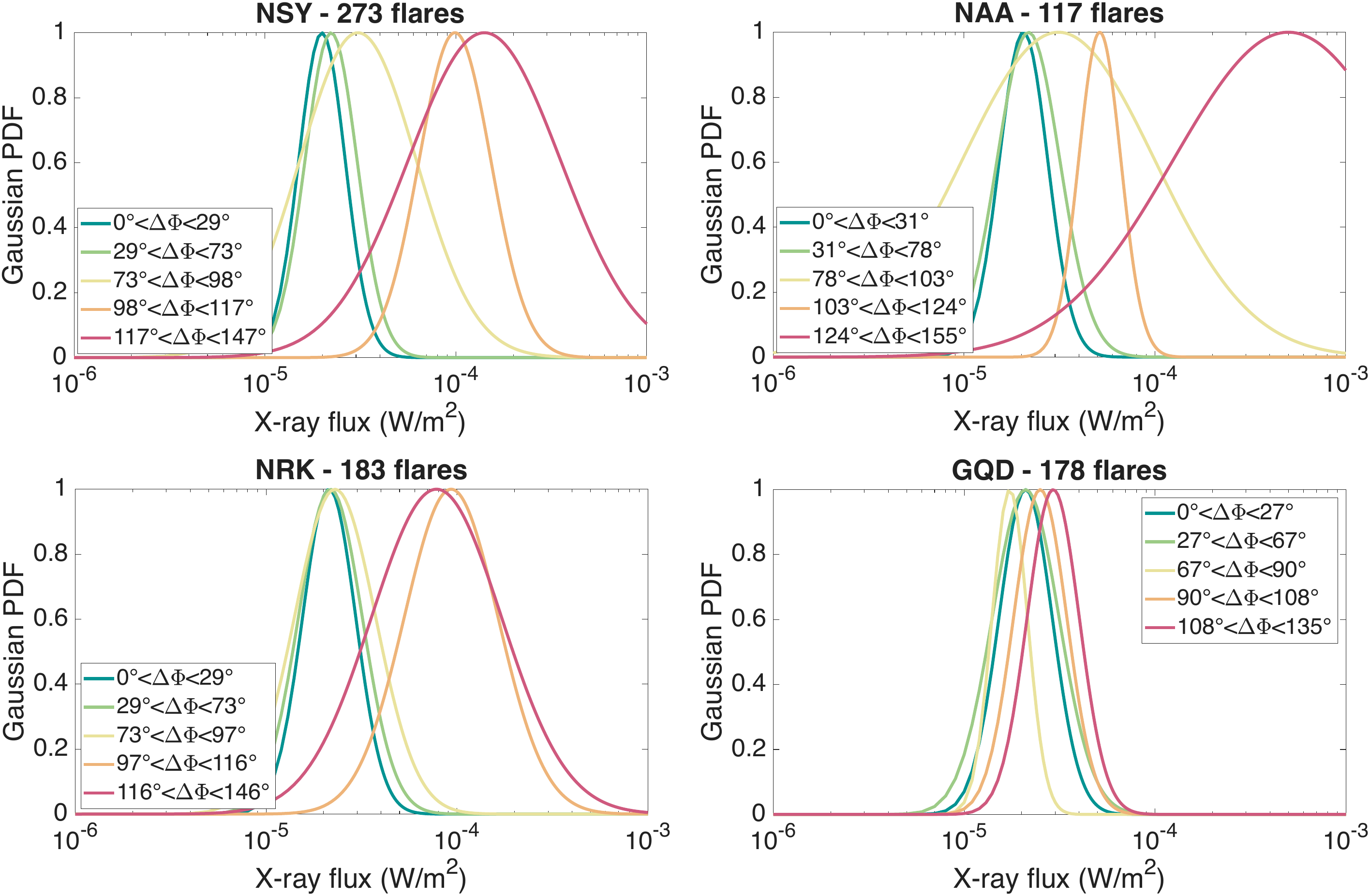}
    \caption{Probability density functions for the four transmitters monitored in Nançay. The PDFs shown in the bottom left panel for NRK were already present in Fig. \ref{fig-fluxNAA}}
    \label{fig-pdfNC}
\end{figure}

PDF are computed for every transmitter monitored in Nançay (Fig. \ref{fig-pdfNC}). For most of them, PDFs behave as expected: larger $\Delta \Phi$ corresponds to higher X-flux. However, GQD phase changes do not strongly depend on the solar flare X-ray flux: the probability densities are almost similar for all $\Delta \Phi$ values. Hence, transmitters behaving like GQD should not be included in the X-flux estimate. The phase behaviour is not an intrinsic property of the transmitter and must be verified for each propagation path. In Nançay, the flux estimation no longer relies on GQD data. 

The quality of the PDF is limited by the dataset used for the training: the small number of flares above X5 limits their detectability. Indeed, the PDFs built on NRK data (bottom left panel, Fig. \ref{fig-pdfNC} show considerable variation if $\Delta \Phi$ is above 116°: the 95\% confidence interval is thus [M1.7, X3.6], due to there being only 13 X-class flares in the training period. Similarly, this confidence interval is [X1.0, X4.2] (resp. [M2.6, X9.3]) for NSY (resp. NAA), relying on 15 (resp. 5) X-class flares. This can be improved by generating synthetic data obtained by paring a chemistry scheme and a VLF propagation model. We will discuss this point in more details in Section \ref{sec-discussion}.

\subsection{PDFs from multiple transmitters}

\begin{figure}
    \centering
    \includegraphics[width=0.5\linewidth]{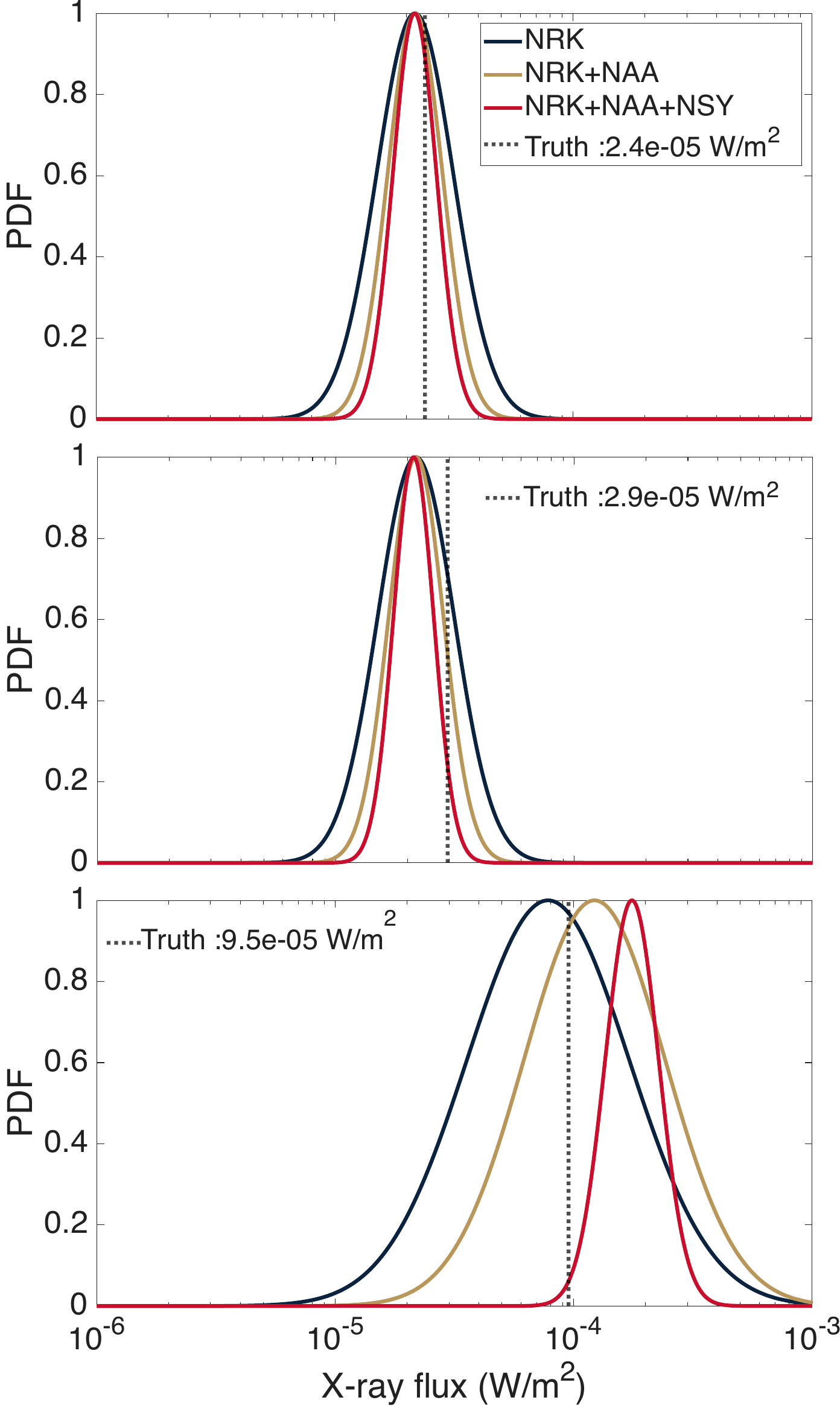}
    \caption{Estimation of the X-flux using one, two or three transmitters for: top panel, M2.9 flare (14.32 UT on 2024/11/05); middle panel, M9.5 flare  (12.11 UT on 2024/11/10); bottom panel, M2.4 flare  (13.14 UT on 2024/12/07). The posterior probability density functions obtained with Equation \ref{eq-proba} are indicated, combining information from one (black), two (brown) or three (red) transmitters.}
    \label{fig-resnovdec}
\end{figure}

The PDF were built on data from November 2023 to October 2024, and the validation sample consists of data from November and December 2024. During those two months, two M2 and one M9 flares were detected by NAA, NSY and NRK. For each flare, the flux is evaluated using NRK only, NRK + NAA or the three transmitters (Fig. \ref{fig-resnovdec}). \\

As the transmitter number increases, the confidence interval reduces. In all but one case, the true X-flux is included within the $2\sigma$ confidence interval (shaded area) deduced from the VLF measurements. However, the estimate does not improve significantly with the combination of transmitters: for the M9 flare, the best flux estimate is even found using a single transmitter. This is partly due to the limited number of samples of intense flares available for training, which affects the phase grid discretization and the training process itself. Yet, maintaining several transmitters in the evaluation reduces the impact of a transmitter's failure, reduces the number of false positives (Section \ref{sec-discussion}), and provides a comprehensive overview of electron density. 

\section{Electron density estimate}\label{sec-necomp}
Computing the D-region electron density is necessary to estimate the solar flare impact on HF absorption, and is achieved through the \textit{Longwave Mode Propagator} \citep[LMP, ][]{gasdia_new_2021}. The electron density profile is described by the Wait profile \citep{wait1964characteristics,thomson_experimental_1993}:

\begin{equation}\label{eq-NE}
    Ne = 1.43\times10^{13} \exp\left(-0.15 h' + (\beta - 0.15)(z - h')\right)
\end{equation}
$h’$ is the effective ionosphere height, $\beta$ is the inverse of the typical gradient length and $z$ is the altitude in the ionosphere, with $z = 0$ on the ground. \\

\begin{figure}
    \centering
    \includegraphics[width=\linewidth]{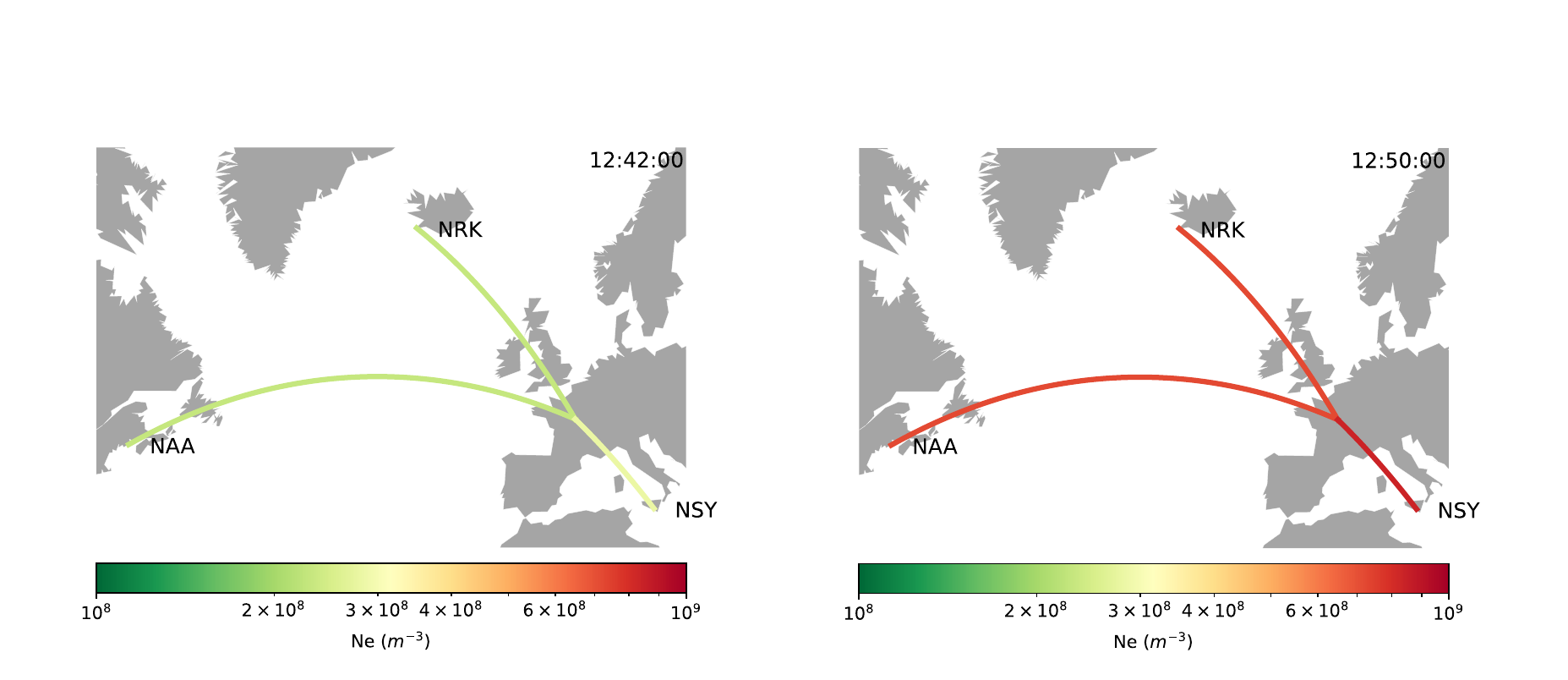}
    \caption{Maps representing the electron density at 70 km in quiet times (left) or at the peak of the M7 flare on 2024/07/28 (right).}
    \label{fig-nemaps}
\end{figure}

For each transmitter, amplitude and phase lookup tables are generated for the range of ($h’$, $\beta$) expected during quiet times to intense flares \citep[Fig 1]{teysseyre2025effect}. Given a measured amplitude and phase, the values of ($h’$, $\beta$) minimising the difference between modelled and measured amplitude and phase are sought, according to the method in \cite{teysseyre2025effect}. Electron densities are thus deduced for each time step of an observed profile, from which maps of electron density along each path are generated (Fig. \ref{fig-nemaps}). These maps show each propagation path colour-coded for its electron density at 70 km in altitude. Note that in practice, the colour-code adapts to high electron densities without saturating at red as the right panel of Figure \ref{fig-nemaps}. 


\section{The \textit{vlf4ions} Python package}\label{sec-package}

Based on the methods presented in previous Sections, we designed a Python package \textit{vlf4ions}. While designed for magnetic loop antennas, it is also suitable for VLF electric-antenna types. Open-source access is provided for the source code (version 1.3.6 at the time of this paper) and its documentation. \\

\begin{figure}
    \centering
    \includegraphics[width=\linewidth]{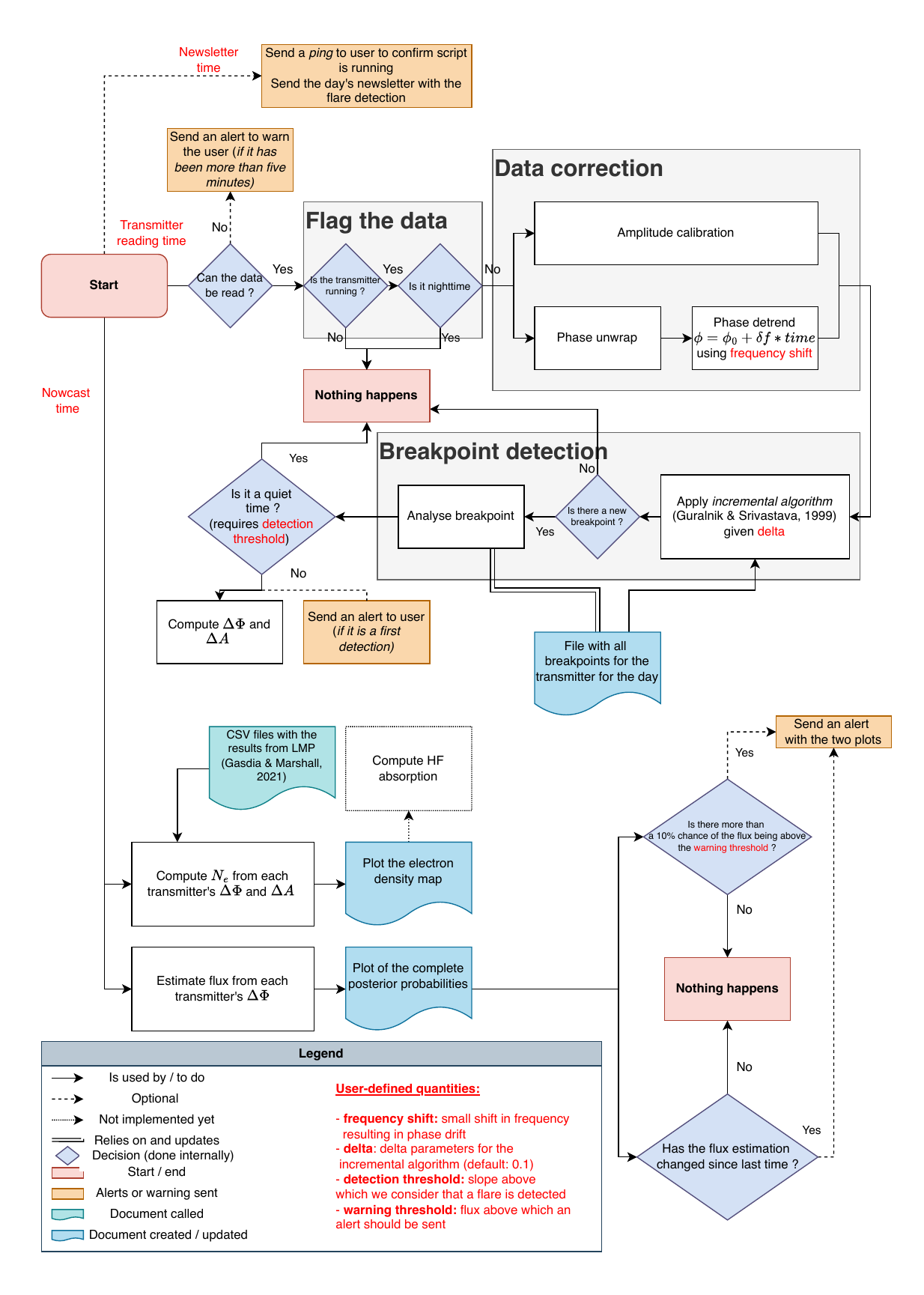}
    \caption{Diagram of the main steps of the real-time data correction, flare detection and electron density computation from the \textit{vlf4ions} package. The different methods and results are detailed in the next sections.}
    \label{fig-globaldiagram}
\end{figure}

Diagram \ref{fig-globaldiagram} depicts the step-by-step functionality of this library’s methods. VLF data analysis using the methods from earlier sections yields solar flare detection, solar X-ray flux and D-region electron density estimates. The `Data correction' and `Flag the data' blocks correspond to Section \ref{sec-datacor}, `Breakpoint detection' is described in Section \ref{sec-detection} and the `Nowcast time' branch is detailed in Sections \ref{sec-fluxestimation} and \ref{sec-necomp}. Different branches are called at given times (specified by the user) to do separate tasks (see Fig. \ref{fig-globaldiagram} from 'Start'):

\begin{itemize}
    \item Newsletter time: A short email is sent to the VLF receiver maintainer as proof that the script is still running. A newsletter summarising daily flare detections (start time, end time and estimated peak flux) is also sent.
    \item Transmitter reading time: Each transmitter’s data is read to detect flares (with a frequency given by $t_r$).
    \item Nowcast time: Evaluations of the solar X-ray flux and electron density profiles along each monitored propagation path are produced at the nowcast time (with a frequency given by $t_r$).
\end{itemize}

 Every necessary parameter ($\delta$, $t_r$, the threshold for the flags and the thresholds for the flare detection) are given by the user in a separate script. Alerts are sent at each flare detection to various recipients. Each alert includes the solar X-ray flux estimation and the electron density profile along monitored paths. \textit{vlf4ions} users can customise these alerts, as detailed in the documentation. Maintenance warnings are sent if the receiver is off so that the instrument maintainer knows to restart the instrument, and newsletters recapitulating each flare detection are crafted at the end of each day.

\section{Improving the detection lead-time}\label{sec-improving}

\begin{figure}
    \centering
    \includegraphics[width=\linewidth]{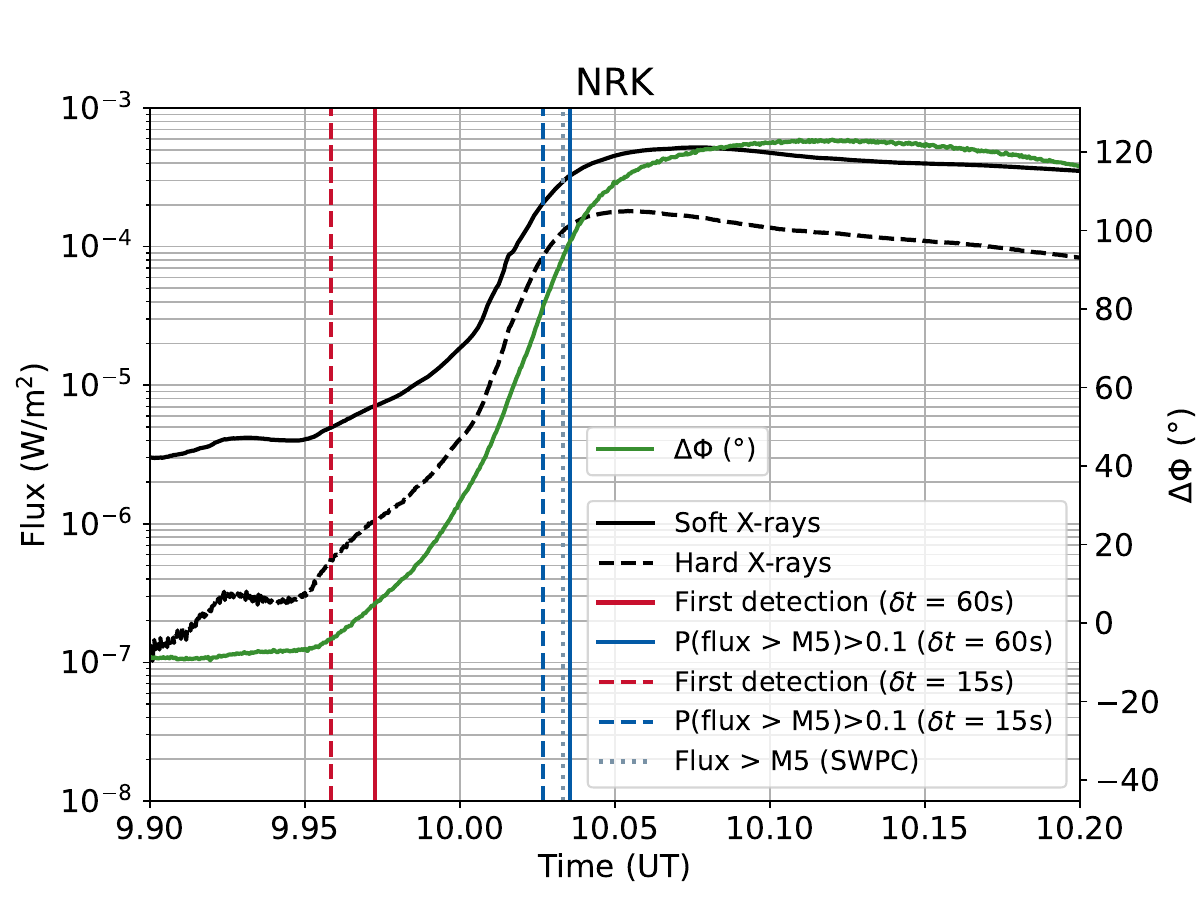}
    \caption{Detection of the X5 flare occuring on 2025/11/11. The soft and hard X-rays fluxes are represented in black (full and dashed lines respectively) and the phase from NRK in green. The red vertical lines represent the first detections of this flare, and the blue vertical lines the moment at which the flux estimation reaches an M5 threshold. Full lines show the time detection with $\delta t$ = 60s and dashed lines those with $\delta t$ = 15s. The blue dotted line is the time of the issue of the SWPC alert for the M5 threshold.}
    \label{fig-11Nov}
\end{figure}

The detection time is improved by decreasing the time between two data reading, $\delta t$: instead of 60 s, data can be read every 15 s which maximises the chance of detecting a flare early. For illustration, Figure \ref{fig-11Nov} shows the difference in detection times when $\delta t$ is 1 min or 15 s, using the data from three transmitters recorded in Nançay. Lowering $\delta t$ to 15 s improves the detection time by 51 s, and the detection of the M5 threshold by 31 s. \\

Similarly, lowering the probability threshold at which the flare alerts are sent (Figure \ref{fig-globaldiagram}) increases the chance to detect all flares and improves the lead-time. However, there is a trade-off with the false positive times, defined here as times when the flux estimation was above M5 (with a given probability threshold, $P_t$). As an illustration, Table \ref{tab-rev2} below presents the percentage of time with false positives, for the months of November and December 2024. The only days considered for this study were those when all three transmitters were working, and without flares above M5, which represents 19 days in total. Neither changing $\delta t$ nor $P_t$ significantly impact the false positive time, though some variations are observed with the parameters. For the envisioned usage of this package, it is thus advised to choose $\delta t$ = 15 s and $P_t$ = 0.1.

\begin{table}[]
    \centering
    \begin{tabular}{l|l|l}
     & $\delta t = 15s$ & $\delta t = 60 s$ \\
    $P_t =$ 0.1 & 2.5 & 2.2 \\
    $P_t =$ 0.25 & 2.5 & 2.2 \\
    $P_t =$ 0.5 & 2.2 & 1.7 \\
    \end{tabular}
    \caption{Percentage of false positive time during selected days in November and December 2024 (see text for more details), depending on the probability threshold and $\delta t$. A probability threshold of 0.1 signifies that a flare alert is sent of P(flux $>$ M5) $>$ 0.1.}
    \label{tab-rev2}
\end{table}

\section{Discussion and conclusion}\label{sec-discussion}

\textit{vlf4ions} is a Python package for VLF receiver owners that analyses VLF data in real-time. An incremental algorithm detects solar flares in real-time by seeking trend breaks in VLF phase data. 82.7\% of M and X flares are detected within one fourth of their rise times. Solar X-ray flux estimate is performed using combined probability density functions derived from the phase variations of several transmitters. The computation is accurate within a factor of two with the current PDF built, and as assessed on our validation dataset. Finally, the D-region’s electron density profile is estimated by comparing VLF changes to those predicted by models like LMP. \\

Note that the method presented here detects changes in the D-region ionisation. Those were attributed to solar flares in the context of this work. However, any change in D-region ionisation will be detected if it is intense enough to affect the VLF phase data. This also implies that flares that do not greatly impact the D-region will be missed. This includes both faint flares and flares occurring close to sunrise or sunset, which do not affect the D-region very much probably due to their very low solar zenith angle. \\

This approaches complements the flare detection systems already in place. Compared to the previous suggestions of a VLF ground-based flare detection systems, the main improvement is the reliance solely on the instrument's short term data. Thus, access to GOES X-ray flux or to past days of data is not required, which increases the resilience of these methods. We also provide estimates of the D-region electron density and the Sun's X-ray flux. This method is adaptable to other VLF receivers; the steps to achieve this are presented in the package's documentation. \\

The most recent versions of this package (v1.3.4 to v1.3.6) have been running in Nançay since 2025/07/29. Previous versions were tested since April 2025. A computationally efficient method is imperative to work in real-time on the instruments’ computers, without disturbing the data recording. The data analysis for a single transmitter is completed in less than one second. Both flux estimate and electron density computations require a maximum of two seconds. Consequently, the method demonstrates suitability for real-time application. \\

Various improvements to the flare detection and X-flux estimate are still possible. The analysis still relies on past data for the flux estimate. An alternative solution for deriving the different probability density functions is to employ synthetic data provided by a chemistry model. Indeed, electron and ion profiles can be derived from various chemistry models, for example WACCM-D \citep{verronen2016waccm}, GPI \citep{glukhov1992relaxation, lehtinen_bluejets} or other chemistry schemes \citep[e.g.][]{mitrarowe_1972}. These models are able to reproduce the ions species and electrons for different forcing sources. The densities can then be inputted into a propagation code for different forcing conditions to estimate the VLF wave time profile both in amplitude and phase. This would have a second crucial improvement: it would allow a finer discretization in $\Delta \Phi$. It would also provide an extension of the probability density functions towards C-flares, which were not included here in the pdf as measuring and validating their $\Delta \Phi$ values is very time-consuming. It thus would be unrealistic to perform this task every time this method is applied to a new antenna. In addition, it would extend the flux estimation to strong X-flares, which were not included in the pdf due to the lack of data. Finally, this flux estimate running in real-time depends only on the values of $\Delta \Phi$ for each transmitter. Thus, this method works similarly in post-processing. \\

Phase measurement was preferred in the present version of the code, as it presents a simpler response to flares compared to amplitude, which exhibits more diverse responses \citep[e.g.][]{Briand22}. Yet, inverted or through profiles also carry information on the flare strength. Seeking negative slopes in the amplitude data may offer a different solution for flare detection and flux estimation. Moreover, relying on $\Delta A$ as well as $\Delta \Phi$ for the flux estimation would reduce the errors and produce estimates with better confidence. There is significant work to achieve this, since the amplitude response to solar flares is more diverse than the phase's. This is thus planned for a future version of the tool. \\

Phase measurements also require careful correction. An alternative to them would be the use of polarisation parameters \citep{gross2018polarization}. Indeed, three out of the four polarisation parameters are unaffected by the phase drifts. Using the tilt angle as well as major and minor axis would provide three quantities on which to base the flare detection and subsequent steps. \\

In this implementation of the electron density computation, the ionosphere was assumed to follow the Wait profile, as the most commonly-used model for the D-region. It also guarantees a fast computation-time and an easy comparison with the existing literature \citep[e.g][]{thomson_daytime_2011, thomson_ionospheric_2022}. This two-parameter model however misses some fine features of the actual electron density profile during flares \citep[e.g.][]{gasdia2023method}. This could be improved in later versions of the \textit{vlf4ions} package by relying on pre-computed profiles from GPI-4 or GPI-5 instead. \\

Since the slopes in the phase data are detected, a short-term forecast of a few minutes may also be produced by extrapolation of the phase slope on short timescales. Thus, alerts could be sent sooner, although this may result in more false positives. This will be studied and implemented in a future version of the package. \\

The performance of this system is quantified using the alerts issued by the SWPC if the X-ray flux reaches M5 (available at \url{https://www.swpc.noaa.gov/products/alerts-watches-and-warnings}). It should be emphasised that these alerts are not sent at the beginning of the flares and only concern flares above M5. On average, there are 11.3 minutes between the flare alerts obtained from VLF data and the alerts issued by the SWPC (the median is 5.2 min, for 39 flares). However, at this point, alerts from VLF data do not distinguish between weak M flares and stronger ones. Though there is not enough data yet to systematically compare the alerts at the M5 level with our system and from GOES data, it is presented for a specific example. For the flare in Figure \ref{fig-11Nov}, the M5 alert was issued first with VLF data with $\delta t = 15 s$. Then, 23 s later (resp. 36 s), it was sent by SWPC (resp. our system, with $\delta t = 60s$). The detection of M5 flares is thus equivalent between the different systems. Note that the ionosphere reacts to the flare with a small delay \citep[e.g.][]{appleton1953note, basak2013effective, Briand22}. However, this delay decreases with flare strength. Indeed, VLF data is sensitive to hard X-rays, as soft X-rays do not penetrate in the D-region \citep[][p. 186, Figure 25]{ratcliffe1960}. The stronger the flare, the earlier the hard X-ray peak occurs compared to soft X-rays. The VLF phase will thus react faster for strong flares, providing earlier alerts. \\ 

The main improvement to this system will be the computation of the HF absorption caused by the flares. Since the electron density is already estimated, this could be achieved through a different chemistry model or by implementing tabulated ion profiles and collision frequencies. This would allow a single receiver to produce HF absorption estimates over an entire continent or ocean, without relying on satellite data. \\

\begin{acknowledgements} 
\end{acknowledgements}

\begin{funding}
      The CNRS supported the present work through its PNST/ATST program, as well as the CSAA. The CNES also supported this work through its support to Space Weather programs. 
\end{funding}

\begin{conflictofinterest}
    The authors declare no Conflict of Interest.
\end{conflictofinterest}

\begin{dataavailability}
The open-source code and documentation for the \textit{vlf4ions} library are found at \url{https://codeberg.org/pteysseyre/vlf4ions_library}. At the time of the paper, the library's version is 1.3.6.  The VLF data from the AWESOME receiver in Nançay is accessible by emailing the authors.

\end{dataavailability}

\bibliography{references}
   


\begin{appendix} 

\end{appendix}

\end{document}